\documentclass[nologo,11pt,a4paper]{ETHpaper}
\usepackage{appendix}
\usepackage{multirow}
\usepackage{amssymb,amsmath} % Mathematical Symbols
\usepackage{graphicx} % Figures
\usepackage[sort&compress]{natbib} % Bibliography
 \usepackage[format=plain,justification=justified,hang]{caption}
\usepackage[utf8]{inputenc}
 \usepackage{cancel}
 \usepackage{pst-pdf} \usepackage{pst-node}
\usepackage{xcolor} \usepackage{soul} \usepackage[hang]{subfigure}
\usepackage{hyperref}
\usepackage{array}
\begin{document}

\newcommand{\bbox}[1]{\mbox{\boldmath $#1$}}
\newcommand{\mean}[1]{\left\langle #1 \right\rangle}
\newcommand{\abs}[1]{\left| #1 \right|}

\newcommand{\avg}[1]{\left< #1 \right>} % for average
\let\underdot=\d % rename builtin command \d{} to \underdot{}
\renewcommand{\d}[2]{\frac{d #1}{d #2}} % for derivatives
\newcommand{\pd}[2]{\frac{\partial #1}{\partial
    #2}} % for partial derivatives
\newcommand{\pdd}[2]{\frac{\partial^2 #1}{\partial #2^2}} % for double
% partial derivatives
\newcommand{\mpd}[3]{\frac{\partial^2 #1}{\partial #2 \partial #3}} %for
% mixed partial derivative

\title{Quantifying the effects of social influence}
\titlealternative{Quantifying the effects of social influence}
\author{Pavlin Mavrodiev, Claudio J. Tessone, Frank Schweitzer}
\authoralternative{P. Mavrodiev, C. J. Tessone, F. Schweitzer}
\address{Chair of System Design, ETH Zurich, Weinbergstrasse 58, 8092
   Zurich, Switzerland}

\www{\url{http://www.sg.ethz.ch}}

\makeframing
\maketitle
\begin{abstract}
How do humans respond to indirect social influence when making decisions?
We analysed an experiment where subjects had to repeatedly
guess the correct answer to
factual questions, while having only aggregated information about the answers
of others.
While the response of humans to aggregated information is a widely
observed phenomenon, it has not been investigated quantitatively, in a controlled
setting.
We found that the adjustment of individual guesses depends
\emph{linearly} on the distance to the mean of all guesses. This is a
remarkable, and yet surprisingly simple, statistical regularity.
 It holds across all questions analysed, even though the correct answers
 differ in several orders of magnitude.
Our finding supports the assumption that individual diversity
does not affect the response to indirect social influence.
It also complements previous results on the nonlinear response in
information-rich scenarios. We argue that the nature of the response to social
influence crucially changes with the level of information aggregation.
This insight contributes to the empirical foundation of models for
collective decisions under social influence.
\end{abstract}
\date{\today}

To what extent are the opinions we hold about subjective matters the result of our own
considerations or a reflection of the opinions of others? Even
though we would like to believe the former, in most real-life situations
individual opinions are highly
interdependent. They are, directly or indirectly, influenced by cultural
norms, mass media and interactions in social networks. The combined
effects of these influences is known as \emph{social influence} --
individuals acting in accordance to the beliefs and expectations of
others \citep{Kahan1997}. Social influence can be
categorised as \emph{direct} or \emph{indirect}. The former is the
result of one individual directly affecting the opinion of another,
typically through coercion or persuasion. The latter is a more subtle
psychological process and takes place
when one's opinion and behaviour is influenced by the availability
of information about others' actions. Our main focus in this paper is on
the second form, therefore we regard social influence as implicitly indirect.

Social influence can be readily
observed in common collective decision processes, e.g. political
polls \citep{Diana1992}, panic stampedes \citep{Helbing2000}, stock markets
\citep{Hirshleifer2003}, cultural markets \citep{Salganik2006}, or aid campaigns \citep{Schweitzer2008}. Some of these
collective decisions can trap a population in a suboptimal
state, for example a financial bubble due to financial actors' herding behaviour
\citep{Prechter2001}. Alternatively, they may steer a system into positive
directions, such as increased tax compliance rates
\citep{Wenzel2005}. However, understanding how such collective decisions
are formed, evaluating their benefit for the population, and even directing
their outcomes, is conditional
on quantifying how people perceive and respond to social influence.

Theoretical work in this field requires to specify a
social structure together with mechanisms by which influence
exerted by that social structure is internalised by the
individuals \citep{Castellano2009}. Typically, it is considered that
individuals form opinions in an interaction network (defined in terms of
their social acquaintances) in which they are
subject to complex inter-personal influences.

As early as 1956, French postulated a theory of \emph{social power}, in which
social structure is represented as an explicit interaction
network \citep{French1956}. An individual adopts an opinion that equals the
mean of his own opinion and those he interacts with. Assuming that
knowledge about the opinion of others
is available, the theory predicts that well-connected populations invariably
reach consensus.

Later, social psychologists and mathematicians have extended and built
upon French's social power theory. Prominent works account for
weighted averaging of others'
opinions \citep{Friedkin1986},  probability distribution of
opinions \citep{DeGroot1974}, and importance of positioning in the
interaction network \citep{Friedkin1997}. In particular, Latan\'{e} made a notable
quantitative contribution with his \emph{social impact}
theory \citep{Latane1981}, which showed via empirical evidence that the
fraction of individuals conforming to a group opinion is a power function of the
group size (with exponent
less than 1). Recent research has also shown how the identification
of an individual with a group affects the final distribution of opinions
\citep{Groeber2009}. In most models based on interaction
networks, it is usually found that individuals respond in a highly non-linear
manner, e.g.~opinion fragmentation, due to the complexities involved in
inter-personal influences \citep{Hegselmann2002}.

In this paper, we contribute to these theoretical investigations by
analysing a decision-making experiment based on aggregate
information instead of on explicit interaction networks. Our approach
assumes that in some decision-making scenarios it
is not always possible to have full information about others'
opinions. Instead, only some sort of aggregated representation of all
opinions is available, which arguably provides less information. For example,
individual compliance to social norms has been shown to depend on
knowing the average compliance rate in the population
\citep{Groeber2010}. Other examples include book purchases being influenced by best-sellers lists that are typically compiled from average book store
sales \citep{Bikhchandani1998}, or recommender systems offering buyers
products whose quality has been estimated as the average of all ratings
\citep{Zhang2009}. We are, therefore, interested in evaluating
whether individuals react differently when subjected to
 limited information compared to the non-linear response with full
information.

Quantification of human responses to aggregated
information is scarce. We present empirical evidence of how
individuals react to it in a controlled environment. The empirical study
we analyse was conducted by \citet{Lorenz2011}. In  this experiment
individuals were asked to guess the correct answer to six quantitative
questions with an objective answer (such as ``What is the border length
between Switzerland and Italy?'') repeatedly over five experimental
rounds (see Table \ref{tab:exp-setup}). Subjects were assigned to three
different treatments in which they had (i) no information
 about others' guesses during all rounds, (ii) the mean of
 all guesses in the  previous round or (iii) full information about others'
 estimates. Here, we focus on (ii), and report
 a statistically  significant linear dependence between the change in
 one's estimate and the distance of the  previous estimate from the mean.

\section*{Results}

We analyse the following set-up: a set of $N$
subjects were asked six quantitative questions with a clearly defined
objective truth. Individuals did not know {\em a priori} the true answers, and
thus could only provide a guess. Each question was repeated for five consecutive
rounds. At the end of each round, the subjects were presented with either
some
or no information about others' guesses, after which they could revise
their own estimate. Let $x_{i}(t)$ be the guess of individual $i \in
[1,N]$ at round $t \in
[1,5]$ for a particular question. The arithmetic average of all $N$ individuals at time $t$ is then denoted as
$\overline{x}(t)$. In the aggregate regime subjects are presented with
$\overline{x}(t)$ at the end of round $t$ before making their next guess
$x_{i}(t+1)$. We study how the change in one's opinion,
$\Delta x_{i}(t) = x_{i}(t)-x_{i}(t-1)$, is related to its the distance from the mean
in the previous time step $\overline{x}(t-1)-x_{i}(t-1)$.
From the experimental
data \citep{Lorenz2011supp},
we can calculate $\Delta x_{i}(t)$ and $\overline{x}(t-1)-x_{i}(t-1)$
across all rounds, subjects, questions and sessions.

\begin{figure}[hptb]
\centering
\includegraphics[scale=0.2]{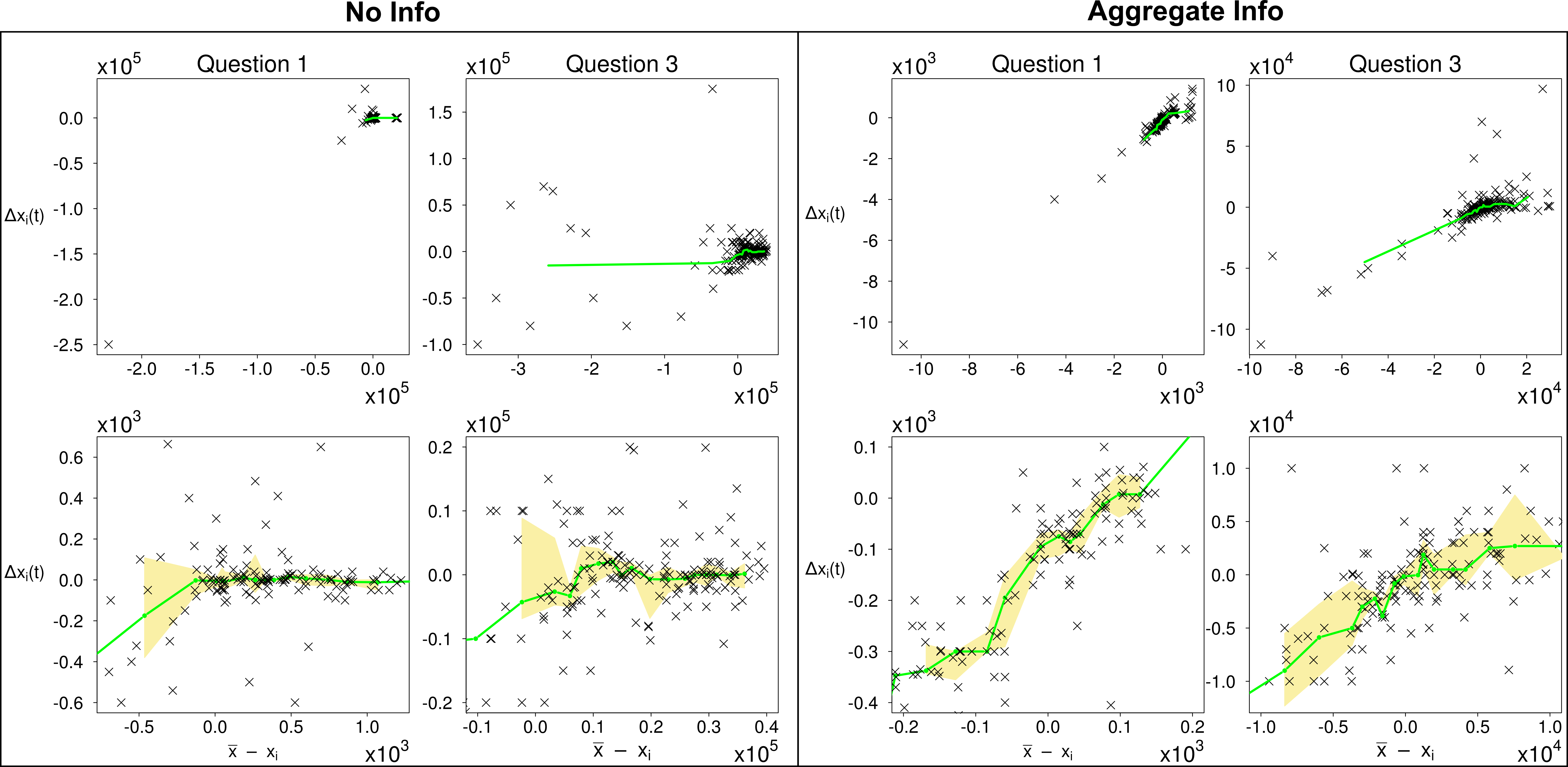}
\caption{\textbf{Scatter plots for questions 1 (first and third column)
    and 3 (second and fourth column).} The green lines show median smoothing: the $x$-axis has been
  split into equally sized bins of size 10 (arbitrary), and the medians in
  each bin are plotted. The bottom row shows median smoothing with
  shaded areas corresponsing to error bars between the first and third quartile of each
  bin. Note the scaling of the $x$- and $y$-axis.  }
\label{fig:questions-main}
\end{figure}

At the finest
granularity of the data, there are $N=12$ subjects answering a given question for a
given information condition over five rounds. In total, one would have
$12 \times 4 = 48$ data points. Considering, however, that each question
was asked four times at a given information condition (see Table
\ref{tab:exp-setup}), we pool these responses together to produce $48 \times 4
= 192$ samples per information condition and per question. In Figure
\ref{fig:questions-main}, we have plotted typical $\Delta x_{i}(t)$
vs. $\overline{x}(t-1)-x_{i}(t-1)$ for two questions. The left column
shows that in the no information regime there is no particular dependency
between the distance to the average and the ensuing adjustment of one's
guess. In contrast, there is a positive linear relation in the aggregate
information regime.

We formalise this qualitative argument by the following linear regression model.
\begin{equation}
\label{eq:linear-model}
\Delta x_{i}(t) = \beta_{0} + \beta_{1}
(\overline{x}(t-1) - x_{i}(t-1)) + \epsilon_{i}(t),
\end{equation}
with the associated null hypothesis
$\mathcal{H}_{0}: \beta_{1} = 0$, and two-sided alternative
$\mathcal{H}_{1}: \beta_{1} \neq 0$.

Due to the experimental set-up, in particular the nature of the
questions, subjects did not have a solid idea about the true
answers. However, the questions were not too hard to prevent educated guesses about the approximate order of magnitude. \citet{Lorenz2011} note that the initial opinion distribution for each
question is right-skewed -- a majority of estimates are low and a
minority fall on a fat right tail. Nevertheless, in Methods, we justify using
Eq. \ref{eq:linear-model} to model the aggregate information regime.

It is important to mention that, in principle, regression models, such as
ours, cannot make explicit claims regarding cause and effect. Rather, the primary goal
is to mathematically derive one variable from the other with as high
fidelity as possible. We posit that in the empirical case considered
here, one is able to infer the main causality direction, because the study was
designed with the main purpose of evaluating how social influence affects one's decisions. Therefore, subjects were exposed to social information \textit{prior} to
their decision making. We, therefore, argue that in the aggregate
regime, one of the main causes for an opinion change is knowledge of the mean
(other causes being unobservable factors, such as conviction in own opinion, beliefs about others' expertise, etc.).

Table \ref{tab:regression} shows all results of estimating the linear
model. We focus primarily on the estimation of $\beta_{1}$, as the
constant term, $\beta_{0}$, is heavily influenced by a few outliers, and thus exhibits
large standard errors even when significant. From the reported
$p$-values, we see that the impact of the distance to the mean opinion,
$\overline{x}(t-1)-x_{i}(t-1)$, is highly significant across all questions (with low
rob. std. errors) in explaining one's own opinion change. Furthermore, the size of the effect shows that knowledge of
the mean accounts for a considerable part of the opinion change.

\section*{Discussion}
Our main goal in this paper was to quantify how people respond to social
influence when making decisions. In particular, we focused on a
limited-information scenario, in which individuals possessed the mean of
all opinions. This form of indirect
social influence is prevalent in a wide range of collective decisions,
e.g. norm compliance, product recommendations and
purchases. Quantifying individual human behaviour in such contexts
contributes to understanding such collective decisions.

We used a unique dataset from an experiment in which subjects had to guess the answer to quantitative
questions repeatedly, while knowing the mean of all guesses. We studied
how the change in individual guesses relates to their
distance from the mean. Our analysis shows that a linear model is sufficient
to explain this relationship for all experimental questions, with a
significant and considerable impact. Furthermore, this finding holds for
questions with correct answers that differ by about 10 orders of magnitude.
Therefore, we emphasize that the result is \emph{not} a first-order
approximation of a non-linear regime around a narrow range of
$\overline{x} - x_i$. 

% We started by analysing an experiment where subjects had to guess the
% correct answer to six quantitative questions over five rounds. In the
% beginning of the first round, subjects formed their initial opinions
% without having any hint to the correct answer. At the end
% of each ensuing round, individuals were presented with (i) no information about
% the estimates of others, (ii) the mean of all estimates in the previous
% round, or (iii) the estimates of all previous rounds. We focused on the
% second information regime and studied how the change in one's opinion
% over time is related to its distance from the mean.

Our quantitative insights represent a striking statistical
regularity. Despite individual differences in subjects, e.g.~emotions,
conviction in one's own opinion, beliefs about the competency of others, and
tendency to conform to the group opinion,
the same mathematical relationship underlies the individual reactions to social influence. This suggests that once initial guesses are formed,
diversity among subjects does not play a role in the adjustment of
subsequent estimates.
% It is, thus, consistent
% with similar non-linear regularities found in information-rich scenarios,
% e.g. social impact theory.
Moreover, we argue that the linear nature of the response is due to the level of information
aggregation in the experiment. We believe that the availability of more fine-grained
information, such as allowing group interactions or providing the opinion
distribution, would recover the complex non-linear response found in
most models of social influence.

Our finding also contributes to the design of agent-based
models for collective decisions. Such models play an important role
in testing individual-level interaction mechanisms that lead a
population to favourable collective decisions. While most prominent
models rely on \textit{ad-hoc} assumptions about
individual behaviour (e.g. linear voter model, Schelling's segregation
model), with the increasing
availability of experimental data, there is a
growing interest in basing these assumptions on empirical regularities.
The rule we revealed can, therefore, be used to further model, quantify and
design collective decisions under aggregated information.

\section*{Methods}
\begin{figure}[ht!b]
\centering
\includegraphics[scale=0.2]{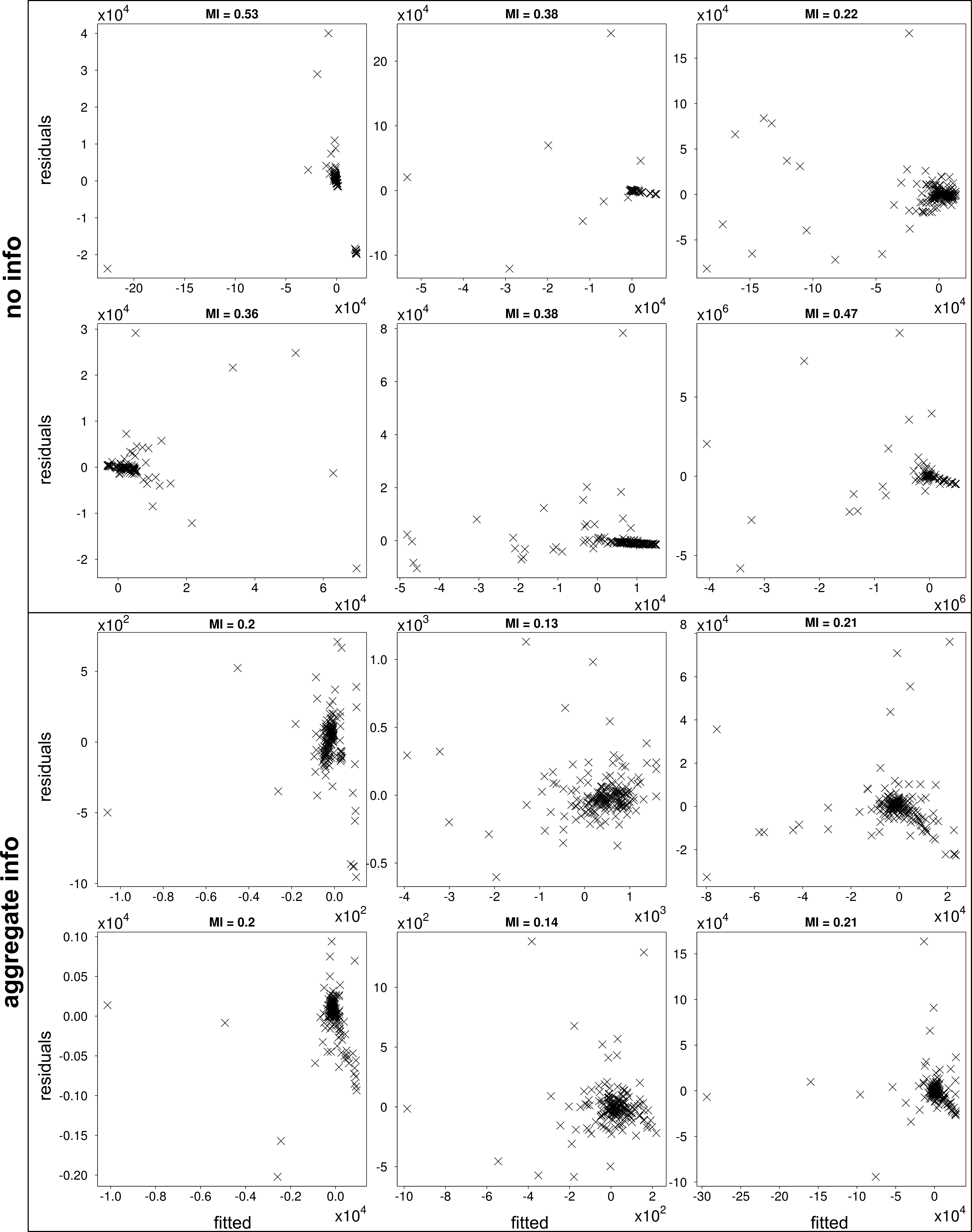}
\caption{\textbf{Residuals vs. fitted values for both information
    conditions and all questions.} The first two rows show
  the no-information condition, while the last two -- the aggregate
  information condition. Questions are numbered from left to right and
  top to bottom. The mutual information (MI) is shown on top of each plot
(see Methods for definition of MI).}
\label{fig:tukey-anscombe}
\end{figure}

The model is estimated by  the method of Ordinary Least Squares (OLS),
which is based to the following assumptions: \textbf{(a)} $E(\epsilon_{i}
| x_{i})=0$ (linear model is correct), \textbf{(b)} $\epsilon_{i} \sim \mathcal{N}(0,\sigma^{2})$
(normality of the error distribution),
\textbf{(c)} $\text{Var}(\epsilon_{i} | x_{i}) =
\sigma^{2}$ (homoscedasticity), and \textbf{(d)}
$E(\epsilon_{i},\epsilon_{j})=0$ (independence of errors).
First, to assess the overall feasibility of the linear model, we plot the
residuals from the OLS estimation of Eq. \ref{eq:linear-model} versus the
fitted values, commonly known as a Tukey-Anscombe plot (Figure
\ref{fig:tukey-anscombe}). A strong trend in the plot is evidence that the linear model
is not suitable, consequently (a) is violated.

For the no-information case, arguably, it is not
reasonable to expect Eq. \ref{eq:linear-model} to be valid %for the no-information
%regime
as subjects did not have access to any
information. Thus, any causal relation between  $\Delta x_{i}(t)$ and
$\overline{x}(t-1)-x_{i}(t-1)$ can be ruled out
\textit{a priori}.

As seen in Figure \ref{fig:tukey-anscombe}, the residuals in
the no information regime do not fluctuate randomly around
the fitted values -- a strong evidence against assumption (a). On the
other hand, comparing with the aggregate
information case, the Tukey-Anscombe plots do not exhibit a visible
dependence between residuals and model fit, thus support assumption (a).

To actually quantify the presence of a trend in Figure \ref{fig:tukey-anscombe},
we compute the  mutual information (MI) between the fitted values
and their residuals. The concept of mutual information originates in
information theory, and, intuitively speaking, measures the amount of
information that two variables share, i.e.~how much knowing one of these
variables reduces uncertainty about the other
\citep[Chap. 2]{Cover2006}. Formally, the mutual information, $I(X,Y)$,
between variables $X$ and $Y$, equals $H(X)+H(Y)-H(X,Y)$, where $H(X)$ is
the information (entropy) in $X$, and $H(X,Y)$ is the joint entropy of $X$
and $Y$. If $X$ and $Y$ are independent then $H(X,Y)=H(X)+H(Y)$, and thus
the mutual information, $I(X,Y)$, equals 0. We also make use of the
inequality $I(X,Y) \le \text{min}\{H(X),H(Y)\}$ to derive the normalisation
$I_{\text{norm}}(X,Y)=I(X,Y) / \text{min}\{H(X),H(Y)\}$. In this way our
MI estimate has an upper bound of 1, which is attained only if $X$ and
$Y$ are identical.

The advantage of computing MI is that it is not
only sensitive to linear correlations, but also to non-linearities that
are not captured in the covariance \citep{Kraskov2004}. The MI
estimations for all questions are shown above each plot in Figure
\ref{fig:tukey-anscombe}. Unsurprisingly, there is stronger
dependency between residuals and fitted values in the no-information
regime, especially where a trend is clearly visible. In contrast, all
questions in the aggregate regime show very low values of MI.

\begin{figure}[hptb]
\centering
\includegraphics[scale=0.2]{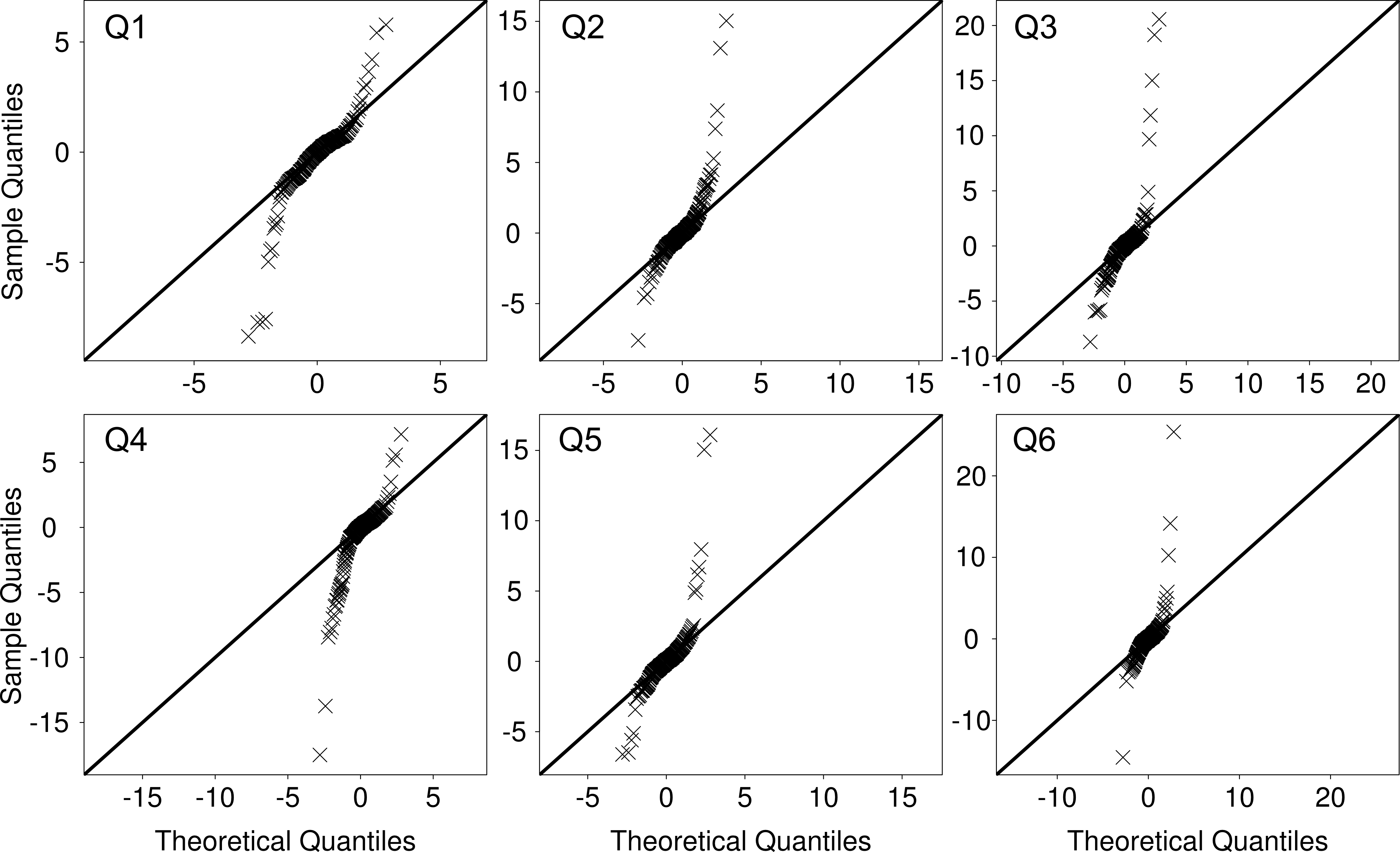}
\caption{\textbf{QQ Plots.} Theoretical quantiles of a normal
  distribution versus sample quantiles for all six questions.
  There are outliers in the data resulting in non-normal
  residuals. Question numbers (Q) are indicated on the top left corner of
each plot. }
\label{fig:qq-plots}
\end{figure}

Second, in Figure \ref{fig:qq-plots} we check normality of errors by plotting the quantiles of the
residual distribution against the quantiles of a normal distribution. The
off-diagonal points in all questions clearly indicate the presence of a few large outliers, as expected for
skewed data. Non-normality of residuals plays no role for the
BLUE (best linear unbiased estimator)
properties of OLS estimators, provided (a) and (c) hold (the
homoscedasticity assumption is evaluated below). However, exact
$t$ and $F$ statistics will be incorrect. Therefore, we make use of the relatively large sample size in all questions to justify the
asymptotic normality property of the OLS estimators
\citep[Chap. 5]{Baltagi2011}. It can be shown that by employing the central limit theorem
and conditional on (a) and (c), OLS produces estimators that are
approximately normal \citep[Chap. 5]{Wooldridge2005}, hence $t$-test
can be carried out in the same way.

Next, we verify the homoscedasticity assumption, (c), of
$\epsilon_{i}(t)$. To this end, we run the Koenker studentised
version of the Breusch-Pagan test \citep{Koenker1981}. This test regresses the squared residuals on the predictor in
Eq. \ref{eq:linear-model} and uses the more widely applied Lagrange
Multiplier (LM) statistics
instead of the $F$-statistics. Although more sophisticated procedures,
e.g. White's test, would account for a non-linear relation between the
residuals and the predictor, we find that the Breusch-Pagan test is
sufficient to detect heteroscedasticity in the data. Table
\ref{tab:heteroscedasticity} shows that the null hypothesis of
homoscedastic error can be rejected with high significance for Questions
1, 2, 4, and 5. The consequence for the OLS method is that the estimated
variance of $\beta_{1}$ will be biased, hence the statistics used to
test hypotheses will be invalid. Furthermore, none of the OLS estimators
will be asymptotically normal. Thus, to account for the presence of
heteroscedasticity, we use robust standard errors.

Finally, the serial correlation in (d) is tested by assuming the following
AR(1) process for the error term
\begin{equation}
\hat{\epsilon_{i}}(t)=\alpha_{0}+\alpha_{1}\hat{\epsilon_{i}}(t-1)+\xi_{i}(t)
\label{eq:ser-cor}
\end{equation}
with $\hat{\epsilon_{i}}$ being the residuals from estimating
Eq.~\ref{eq:linear-model} and $\xi_{i}(t) \sim
\mathcal{N}(0,z^{2})$. One-period lag is sufficient to model error
correlation, given that subjects answered the same question over just 5
rounds. In addition, by excluding the first guess when no information was
available, we have effectively 4 periods. The OLS estimation of
Eq.~\ref{eq:ser-cor} in Table \ref{tab:serial-correlation} indicates that
$\alpha_{1}$ either is not significantly different from 0 (Questions 3, 5 and
6) or has a small effect when significant (Questions 1 and
4). Consequently, inferences based on $t$-tests and $F$-tests can be carried out.

All data analysis was done with R (\url{http://www.r-project.org/},
version 2.15.0). Quantile plots of the residuals were generated with
\verb|rqq| (package \verb|lawstat|,version 2.3). Breusch-Pagan
heteroscedasticity test was implemented by \verb|bptest| (package
\verb|lmstat|, version 0.9-29). Finally to estimate
Eq. \ref{eq:linear-model}, we used the standard \verb|lm| function with
robust standard errors calculated by \verb|hccm| (package \verb|car|,
version 2.0-12). Mutual information was computed with
\verb|multiinformation| (package \verb|infotheo|, version 1.1.0)

\section*{Acknowledgements}
We would like to thank Ingo Scholtes and Antonios Garas for their useful
comments in the early version of this manuscript.

\section*{Author contributions}
P.M. and C.T. designed the analysis. P.M gathered and analysed the data. P.M., C.T., and F.S. wrote the
manuscript.
\section*{Additional Information}
\textbf{Competing financial interests:} The authors declare no competing
financial interests.

\clearpage
\bibliographystyle{sg-bibstyle}
\bibliography{ref}
\clearpage

\begin{table}[h!tb]
\centering
\begin{tabular}{|c|p{1cm}|p{1cm}|p{1cm}|p{1cm}|p{1cm}|p{1cm}|}
\hline
& \textbf{Q1} & \textbf{Q2} & \textbf{Q3} & \textbf{Q4} & \textbf{Q5} & \textbf{Q6} \\ \hline
\multirow{2}{*}{\textbf{no info}} & S1,S4, & S2,S5, &
S3,S6, & S1,S4, & S2,S5, & S3,S6,\\
& S7,S10 &S8,S11 &S9,S12 & S7,S10&S8,S11 & S9,S12 \\ \hline
\multirow{2}{*}{\textbf{agregate info}} & S3,S6, & S1,S4, & S2,S5,  &
S3,S6, & S1,S4, & S2,S5,  \\
& S9,S12 & S7,S10 & S8,S11 & S9,S12 & S7,S10 & S8,S11 \\ \hline
\multirow{2}{*}{\textbf{full info}} & S2,S5, & S3,S6, & S1,S4,  &
S2,S5, & S3,S6, & S1,S4, \\
&S8,S11 &S9,S12 &S7,S10 &S8,S11 &S9,S12 &S7,S10\\ \hline
\end{tabular}
\caption{\textbf{Experimental Setup.} The experiment consisted of 12
  sessions (S) each composed of 12 subjects. In each session, the 12
  subjects had to answer two questions (Q) in the no information, two in the
  aggregate and two in the control condition (see main text), for a total
  of six questions. The order of the questions was randomised across
  sessions. After each of the five rounds subjects were asked the same
  question again and could revise their answers depending on the
  information available to them. In the table, columns
  indicate question number and rows -- information regime. Each cell lists
  the sessions when a given question was asked for a particular
  information regime.}
\label{tab:exp-setup}
\end{table}

\begin{table}[hptb]
\centering
\begin{tabular}{|c|p{1cm}|p{1cm}|p{1cm}|p{1cm}|p{1cm}|p{1cm}|}
\hline
& \textbf{Q1} & \textbf{Q2} & \textbf{Q3} & \textbf{Q4} & \textbf{Q5} & \textbf{Q6} \\ \hline
\textbf{LM statistic} & 0.46 & 11.84 & 0.0037 & 4.607 & 7.5679 & 1.1711 \\ \hline
\textbf{$p$-value} &0.5 & 0.0005 & 0.95 & 0.03 & 0.005 & 0.28 \\ \hline
\textbf{samples} & 192 & 192 & 192 & 192 & 188 & 188 \\ \hline
\end{tabular}
\caption{\textbf{Breusch-Pagan test for heteroscedasticity.} Each column
  corresponds to one of the six questions. Since the linear model has only one regressor the Koenker version of the test has one
  degree of freedom for all questions.}
\label{tab:heteroscedasticity}
\end{table}

\begin{table}[hptb]
\centering
\begin{tabular}{|c|c|c|c|c|c|c|c|}
\hline
& & \textbf{Estimate} & \textbf{Robust std. errors} & \textbf{$t$-value} &
\textbf{$p$-value} & \textbf{N} & \textbf{df} \\ \hline
\multirow{2}{*}{\textbf{Q1}} & $\alpha_{0}$ &-3.6 & 14.02 & -0.26 &
0.79 & \multirow{2}{*}{191} & \multirow{2}{*}{189}\\
& $\alpha_{1}$ & 0.3 & 0.12 & 2.47 & 0.01 & &\multicolumn{1}{|c|}{}\\
 \hline
\multirow{2}{*}{\textbf{Q2}} & $\alpha_{0}$ &0.46 & 12.61 & 0.04 &
0.97 & \multirow{2}{*}{191} & \multirow{2}{*}{189} \\
& $\alpha_{1}$ & -0.19 & 0.1 & -2 & 0.05 & &\multicolumn{1}{|c|}{}  \\
 \hline
\multirow{2}{*}{\textbf{Q3}} & $\alpha_{0}$ &7.2 & 836 & 0.009 &
0.9 & \multirow{2}{*}{191} & \multirow{2}{*}{189} \\
& $\alpha_{1}$ & 0.03 & 0.07 & 0.47 & 0.64 &&\multicolumn{1}{|c|}{}   \\
 \hline
\multirow{2}{*}{\textbf{Q4}} & $\alpha_{0}$ &-1.88 & 22.36 & -0.08 &
0.93 & \multirow{2}{*}{191} & \multirow{2}{*}{189} \\
& $\alpha_{1}$ & 0.05 & 0.16 & 2.14 & 0.03 & &\multicolumn{1}{|c|}{} \\
 \hline
\multirow{2}{*}{\textbf{Q5}} & $\alpha_{0}$ &-0.32 & 14.9 & -0.02 & 0.9 &
\multirow{2}{*}{187} & \multirow{2}{*}{185}
\\
& $\alpha_{1}$ & -0.07 & 0.05 & -1.43 & 0.15 &&\multicolumn{1}{|c|}{} \\
 \hline
\multirow{2}{*}{\textbf{Q6}} & $\alpha_{0}$ &-3.6 & 1388 & -0.003 &
0.99  & \multirow{2}{*}{187} & \multirow{2}{*}{185} \\
& $\alpha_{1}$ & -0.01 & 0.07 & -0.19 & 0.85 &&\multicolumn{1}{|c|}{}   \\
 \hline

\end{tabular}
\caption{\textbf{First-order serial correlation of residuals}. }
\label{tab:serial-correlation}
\end{table}

\begin{table}[hptb]
\centering
\begin{tabular}{|>{\centering\arraybackslash}m{5pt}|>{\centering\arraybackslash}m{10pt}|>{\centering\arraybackslash}m{50pt}|>{\centering\arraybackslash}m{50pt}|>{\centering\arraybackslash}m{50pt}|>{\centering\arraybackslash}m{40pt}|>{\centering\arraybackslash}m{70pt}|>{\centering\arraybackslash}m{10pt}|>{\centering\arraybackslash}m{5pt}|}
\hline
& & \textbf{Estimate} & \textbf{Std. Errors} & \textbf{Robust std. errors} &
\textbf{$t$-value} & \textbf{$p$-value} & \multicolumn{1}{c|}{\textbf{samples}}&\multicolumn{1}{c|}{\textbf{df}} \\ \hline
\multicolumn{1}{|c|}{\multirow{2}{*}{\textbf{Q1}}} & $\beta_{0}$ &-176.46 & 14.98 & 15.55 &
-11.35 & $ < 2.2 \times 10^{-16}$ & \multicolumn{1}{c|}{\multirow{2}{*}{192}}&\multicolumn{1}{c|}{\multirow{2}{*}{190}} \\ \cline{2-7}
\multicolumn{1}{|c|}{}                &
 $\beta_{1}$ & 0.97 & 0.02 & 0.1 & 9.57 & $ < 2.2 \times 10^{-16}$ &
 \multicolumn{1}{c|}{} & \multicolumn{1}{c|}{}  \\
 \hline

\multicolumn{1}{|c|}{\multirow{2}{*}{\textbf{Q2}}} & $\beta_{0}$ & 35.33 & 12.6 & 12.9 &
2.74 & 0.007  & \multicolumn{1}{c|}{\multirow{2}{*}{192}}&\multicolumn{1}{c|}{\multirow{2}{*}{190}} \\ \cline{2-7}
\multicolumn{1}{|c|}{}                        &
 $\beta_{1}$ & 0.27 & 0.05 & 0.09 & 2.89 & 0.004& \multicolumn{1}{c|}{}& \multicolumn{1}{c|}{} \\
 \hline

\multicolumn{1}{|c|}{\multirow{2}{*}{\textbf{Q3}}} & $\beta_{0}$ & -1321.5 & 828.2 & 853 &
-1.55 & 0.12  & \multicolumn{1}{c|}{\multirow{2}{*}{192}}&\multicolumn{1}{c|}{\multirow{2}{*}{190}} \\ \cline{2-7}
\multicolumn{1}{|c|}{}                        &
 $\beta_{1}$ & 0.83 & 0.05 & 0.1 & 6.25 & $2.7 \times 10^{-9}$ &
 \multicolumn{1}{c|}{} &  \multicolumn{1}{c|}{} \\
 \hline

\multicolumn{1}{|c|}{\multirow{2}{*}{\textbf{Q4}}} & $\beta_{0}$ & -146.3 & 23.2 & 23.7 &
-6.2 & $3.8 \times 10^{-9}$  & \multicolumn{1}{c|}{\multirow{2}{*}{192}}&\multicolumn{1}{c|}{\multirow{2}{*}{190}} \\ \cline{2-7}
\multicolumn{1}{|c|}{}                        &
 $\beta_{1}$ & 0.6 & 0.01 & 0.03 & 18.8 & $ < 2.2 \times 10^{-16}$ &
 \multicolumn{1}{c|}{} &  \multicolumn{1}{c|}{} \\
 \hline

\multicolumn{1}{|c|}{\multirow{2}{*}{\textbf{Q5}}} & $\beta_{0}$ & 6.8 & 14.8 & 15.1 &
0.5 & 0.66  & \multicolumn{1}{c|}{\multirow{2}{*}{188}}&\multicolumn{1}{c|}{\multirow{2}{*}{186}} \\ \cline{2-7}
\multicolumn{1}{|c|}{}                        &
 $\beta_{1}$ & 0.4 & 0.04 & 0.1 & 3.72 & 0.0003 & \multicolumn{1}{c|}{} &
  \multicolumn{1}{c|}{}\\
 \hline

\multicolumn{1}{|c|}{\multirow{2}{*}{\textbf{Q6}}} & $\beta_{0}$ & -821 & $10^{3}$ & 1387 &
-0.6 & 0.55  & \multicolumn{1}{c|}{\multirow{2}{*}{188}}&\multicolumn{1}{c|}{\multirow{2}{*}{186}} \\ \cline{2-7}
\multicolumn{1}{|c|}{}                        &
 $\beta_{1}$ & 0.46 & 0.02 & 0.03 & 15.3 & $<2 \times 10^{-16}$ & \multicolumn{1}{c|}{}&\multicolumn{1}{c|}{}\\
 \hline
\end{tabular}
\caption{\textbf{Robust linear regression of Eq. \ref{eq:linear-model}.}
 Uncorrected standard errors are reported for comparison only. Last column
 shows degrees of freedom.}
\label{tab:regression}
\end{table}

\newpage

\end{document}